\documentclass[a4paper]{jpconf}

\usepackage{amssymb,amsmath,graphicx,hyperref}

\numberwithin{equation}{section}

\newcommand{\ve}[1]{\mathbf{#1}}

\begin{document}


\begin{flushright}
\begin{footnotesize}
 MAN/HEP/2015/02\\
 February 2015
\end{footnotesize} 
\end{flushright}
\vspace{-2em}

\makeatletter

\def\ps@headings{%
     \def\@oddfoot{\hfil\thepage\hfil}
     \def\@evenfoot{\hfil\thepage\hfil}
     \let\@oddhead\@empty
     \let\@evenhead\@empty
      \let\@mkboth\markboth
      \let\sectionmark\@gobble
      \let\subsectionmark\@gobble}

\pagestyle{headings}
\addtolength{\footskip}{1.2cm}

\makeatother


\title{Symmetry Improved 2PI Effective Action and the Infrared Divergences of the Standard Model}

\author{\underline{Apostolos Pilaftsis}, \underline{Daniele Teresi}}

\address{Consortium for Fundamental Physics,
  School of Physics and Astronomy, \\ 
  University of Manchester, Manchester M13 9PL, United Kingdom.}

\ead{Apostolos.Pilaftsis@manchester.ac.uk, Daniele.Teresi@manchester.ac.uk}

\begin{abstract}
Resummations of infinite sets of higher-order perturbative contributions are often needed both in thermal field theory and at zero temperature. For instance, the behaviour of the Standard Model (SM) effective potential extrapolated to very high energies is known to be extremely sensitive to higher-order effects. The 2PI effective action provides a systematic approach to consistently perform such resummations. However, one of its major limitations was that its loopwise expansion introduces residual violations of possible global symmetries, thus giving rise to massive Goldstone bosons in the spontaneously broken phase of the theory. We review the recently developed symmetry-improved 2PI formalism for consistently encoding global symmetries in the 2PI approach, and discuss its satisfactory field-theoretical properties. We then apply the formalism to study the infrared divergences of the SM effective potential due to Goldstone bosons, which may affect the stability analyses of the SM. We present quantitative comparisons, for the scalar sector of the SM, with the approximate partial resummation procedure recently developed to address this problem, and show the quantitative discrepancy of the latter with the more complete 2PI approach, thus motivating further studies in this direction.

\end{abstract}

\section{Introduction}
In thermal  field theory, finite-order perturbative expansions
break  down at  high temperatures  and one  needs to
resum higher-order contributions to  deal with this problem. On the other hand, also at zero temperature there are situations where higher-order effects may potentially play an important role, even in a small-coupling regime. For instance, it has recently become well known that the behaviour of the Standard Model (SM) effective potential, extrapolated to very high energies, is extremely sensitive to the physics at the electroweak scale~\cite{Bezrukov:2012sa,Degrassi:2012ry,Buttazzo:2013uya}. Thus, a formalism to incorporate and resum higher-order effects in a consistent manner is highly desirable for of both thermal and non-thermal applications.

A natural  framework to  address  such problems is  the
formalism    introduced   by    Cornwall,    Jackiw   and    Tomboulis
(CJT)~\cite{CJT}.  Its simplest version is known as   the  
Two-Particle-Irreducible (2PI) effective  action. This is an effective action  expressed not only in terms of the field,  but also in terms of dressed  propagators. When one considers a given  truncation to the 2PI effective action, at any given order  of the loopwise expansion, the  2PI effective action
resums, automatically,  an infinite set  of higher-order diagrams  induced by  partially resummed propagators, without the danger of over-counting graphs.

So far, the 2PI formalism has been mainly used in thermal contexts, both at equilibrium~\cite{Blaizot:2000fc,Berges:2004hn,Marko:2013lxa, Alford:2013koa, Fejos:2014ona} and out of equilibrium~\cite{Berges:2004yj, Millington:2012pf, Meistrenko:2013yya, Dev:2014wsa}. One of the major limitations for its application to zero-temperature problems like, for example, the study of the SM effective potential, has been the well-known difficulties of the CJT formalism to incorporate symmetries in a satisfactory way. In particular,  global symmetries are not exactly maintained at a given loop order  of the 2PI expansion,  since they get distorted  by higher-order effects, and this results in violation of the Goldstone theorem  \cite{Goldstone, Goldstone_2}  by higher-order  terms, giving
rise to a massive Goldstone boson in the Spontaneous Symmetry Breaking
(SSB)                   phase                  of                  the
theory~\cite{Baym,AmelinoCamelia,Petropoulos,Lenaghan}.   In the past,
several studies  were presented in the  literature, attempting to
provide       a        satisfactory       solution       to       this
problem~(see references in \cite{PilaftsisTeresi}).

We have recently developed  a new \emph{symmetry-improved CJT  formalism}~\cite{PilaftsisTeresi} that addresses  this long-standing  problem and allows one to consistently use the 2PI effective action to study theories with global symmetries. In  the symmetry-improved 2PI formalism,  the Ward Identities (WIs) associated  with global  symmetries provide additional   constraints  for the extremal solutions of the fields and propagators. This formalism has a number of satisfactory field-theoretical properties, and actually provides an improvement over the standard CJT formulation, in the sense that the behaviour expected for the full theory is recovered already at a low orders of approximation. The symmetry-improved formalism has been also used to study the chiral phase transition~\cite{Mao:2013gva}, and it has recently been extended to higher $n$PI effective actions~\cite{Brown:2015xma}, confirming its consistency also when one goes beyond the 2PI approach.

It has been recently pointed out~\cite{Martin:2013gka} that the SM effective potential suffers from infrared (IR) divergences due to the Goldstone bosons of the electroweak gauge group, which appear at three-loop order. They appear at two loops for the derivative of the potential, i.e. for the minimization condition that fixes the value of the Higgs vacuum expectation value (VEV). This problem needs to be addressed for a two-fold reason. From the conceptual point of view, the effective potential should be well-defined for all values of the Higgs background field $\phi$. On the other hand, from the quantitative point of view, the appearance of these IR divergences formally lowers the order of the involved three- and higher-loop contributions, thus breaking down perturbation theory and potentially giving a significant contribution to the threshold corrections to the VEV. Because of the extreme sensitivity of the effective potential extrapolated at high energies to the matching conditions at the electroweak scale, this could result on quantitative sizable effects on the stability analyses of the SM potential. These issues have been recently addressed by means of an approximate partial resummation procedure, developed in \cite{Martin:2014bca,Elias-Miro:2014pca}. In addition to confirming the disappearance of the IR divergences within our approach, we will show that the more complete 2PI analysis, based on our symmetry-improved formalism, gives quantitatively different results, at least for the scalar sector of the SM. This suggests that even for the full SM there might be quantitative discrepancies between the approximate resummation scheme of~\cite{Martin:2014bca,Elias-Miro:2014pca} and the 2PI approach, thus motivating further studies in this direction~\cite{PT_new}. 

The  layout of  the  paper  is as  follows. In Sections~\ref{sec:CJT} and \ref{sec:SICJT} we review the standard and symmetry-improved 2PI formalisms, respectively. In Section~\ref{sec:HF} we consider the first non-trivial truncation of the effective action, known as the Hartree-Fock (HF) approximation,  for a scalar $\mathbb{O}(2)$ model, and demonstrate some of the satisfactory field-theoretical properties of the formalism. In Section~\ref{sec:sunset} we go beyond this approximation and show that the symmetry-improved approach describes correctly the thresholds of the Higgs and Goldstone self-energies, in agreement with unitarity requirements. Moreover, we demonstrate explicitly how the formalism implicitly resums sets of processes at arbitrarily high order. Finally, in Section~\ref{sec:IR} we use the symmetry-improved 2PI effective action to study the issue of IR divergences in a global $\mathbb{SU}(2) \times \mathbb{U}(1)$ scalar model. We draw our conclusions in Section~\ref{sec:conclusions}.

\section{CJT Effective Action}\label{sec:CJT}
In  this section  we briefly review the CJT formalisms. For concreteness, we consider a $\mathbb{O}(N)$ scalar  model and show that, as opposed to the  1PI formalism, loopwise truncations of the CJT
generating effective action lead to violation of the Goldstone theorem
through higher-order terms.

The model  under consideration  here is the $\mathbb{O}(N)$
scalar model described by the Lagrangian
\begin{equation}
  \label{eq:ONmodel}
\mathcal{L}[\phi]\ =\ \frac{1}{2}\, (\partial_\mu \phi^i)\,(\partial^\mu
\phi^i)\: +\: \frac{m^2}{2}\, (\phi^i)^2\: -\: \frac{\lambda}{4}\,
    (\phi^i)^2\,(\phi^j)^2 \;,
\end{equation}
where $\phi^i =  \big(\phi^1\,,\, \phi^2\,,\, \cdots\,,\, \phi^N\big)$
represents the $\mathbb{O}(N)$ scalar multiplet and summation over the
repeated indices $i,j = 1,2,\dots,N$ is implied.  At zero temperature,
$T=0$, the model  has a SSB phase for $m^2>0$,
with   breaking   pattern:    $\mathbb{O}(N)    \to
\mathbb{O}(N-1)$.   As a consequence  of  the Goldstone  theorem \cite{Goldstone, Goldstone_2}, there are $N-1$ Goldstone bosons and one Higgs particle $H$.
For the  simple case  $N=2$, which we are  considering here, the
field  components~$\phi^{1,2}$  may  be
decomposed as 
\begin{equation}
\phi^H\ \equiv\ \phi^1\ =\ \langle \widehat\phi^1 \rangle\: +\: H \;, \qquad
\phi^G\ \equiv\ \phi^2\ =\ G \;,
\end{equation}
where we have denoted field operators with
a caret.  Here, $G$ is  the Goldstone field,
which is massless at the  minimum of the potential, whereas $H$ is
the Higgs boson, which is in general massive.

\begin{figure}
\centering
\includegraphics[width = 0.8\textwidth]{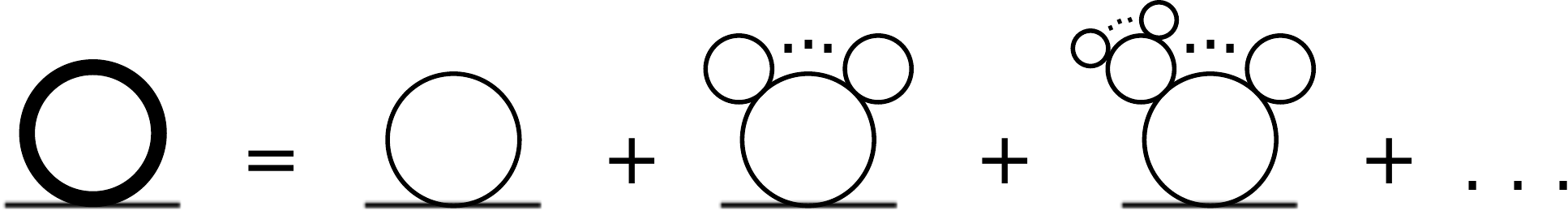}
\caption{\label{fig:daisy} Infinite set of perturbation-theory diagrams implicitly resummed, in the 2PI formalism, by the self-energy on the LHS. Thick lines represent the full propagator, thin lines the tree-level one.}
\end{figure}

The CJT  effective action is the generalization of  the 1PI  one,
where in addition to the local source $J_x$, multi-local sources are introduced.   In its simplest version, the 2PI formalism, we consider one local and one bi-local source,  i.e.~$J_x$ and $K_{xy}$.  Therefore, the
connected generating functional $W[J,K]$ is given by
\begin{equation}
W[J,K]\ =\ - i\, \ln \int \mathcal{D} \phi^i \, \exp\bigg[{i \bigg(S[\phi]
    \: +\: J_x^i \,\phi_x^i \: +\: \frac{1}{2} \,K_{xy}^{ij}\, \phi_x^i
    \,\phi_y^j\bigg)}\bigg] \;, 
\end{equation}
where $S[\phi] = \int_x\, {\cal  L}[\phi ]$ is the classical action. Here and in the  following, repeated spacetime coordinates will denote
integration   with  respect  to   these  coordinates.  The  background  fields
$\phi^i_x$     and    the  connected    propagator $\Delta^{ij}_{xy}$  are  obtained  by  single  and  double  functional
differentiation of $W[J,K]$ with respect to the source $J^i_{x}$:
\begin{equation}
\frac{\delta W[J,K]}{\delta J_x^i}\ \equiv\ \phi_x^i \;,\qquad 
- i\,\frac{\delta W[J,K]}{\delta J_x^i \, \delta J_y^j}\ =\ \langle
\widehat{\phi}^i_x \widehat{\phi}^j_y\rangle - \langle
\widehat{\phi}^i_x\rangle \langle \widehat{\phi}^j_y \rangle\ \equiv\ i
\Delta^{ij}_{xy} \;. 
\end{equation}
We also have
\begin{equation}
\frac{\delta W[J,K]}{\delta K^{ij}_{xy}}\ =\ \frac{1}{2}\, 
\Big( i \Delta^{ij}_{xy} \: +\: \phi^i_x \,\phi^j_y \Big)\, .
\end{equation}
The 2PI effective action $\Gamma[\phi,\Delta]$ is obtained as the
double Legendre transform of $W[J,K]$ with respect to $J$ and $K$:
\begin{equation}
\Gamma[\phi,\Delta]\ =\ W[J,K]\: -\: J_x^i \, \phi_x^i\: -\: 
\frac{1}{2} \, K_{xy}^{ij}\, \Big(i \Delta^{ij}_{xy}\: +\: 
\phi^i_x \phi^j_y\Big) \; . 
\end{equation}
This has the explicit form~\cite{CJT}:
\begin{equation}
  \label{eq:2PI_Gamma}
\Gamma[\phi, \Delta]\ =\ S[\phi]\: -\: 
\frac{i}{2}\, \Tr \Big(\ln \Delta \Big)\: +\: \frac{i}{2}\, 
\Tr \Big(\Delta^{(0)\,-1}\,\Delta\Big)\: -\: i \Gamma^{(\geq 2)} \;,
\end{equation}
where  $\Delta^{(0)\,-1,ij}_{xy} =  \delta^2  S[\phi]/(\delta \phi_x^i\,
\delta  \phi_y^j)$ is  the  inverse tree-level  propagator matrix  and
$\Gamma^{(\geq  2)}$ is the diagrammatic series of all two-  and higher-loop  2PI vacuum diagrams in which the dressed propagator matrix  $\Delta$ is used, instead of the tree-level one.  The
equations of motions (EoMs) are obtained as
\begin{equation}
  \label{eq:2PIextrema}
\frac{\delta \Gamma[\phi, \Delta]}{\delta \phi_x^i}\ =\ 
- J_x^i\: -\:  K_{xy}^{ij} \, \phi_y^j \;,\qquad\qquad
\frac{\delta \Gamma[\phi, \Delta]}{\delta \Delta_{xy}^{ij}}\ =\ -\, 
\frac{i}{2}\, K_{xy}^{ij} \; .
\end{equation}
In  the limit of  vanishing external sources,  the  physical  solution  is  obtained  by  extremizing  the  2PI effective action $\Gamma[\phi,\Delta]$. 
In particular, the EoM for the propagator has the form
\begin{equation}
\Delta^{-1}\ =\ \Delta^{(0)\, -1}[\phi]\: +\: \Pi[\phi, \Delta] \;,
\end{equation}
where  $\Pi[\phi,\Delta]$  is  the   1PI  self-energy,  in  which  the
propagator lines are given  by the dressed propagator matrix $\Delta$.
When a given truncation of $\Gamma[\phi,\Delta]$ is explicitly considered, this EoM implicitly resums an infinite set of perturbation-theory Feynman graphs, as shown in Fig.~\ref{fig:daisy}.

By considering the $\mathbb{O}(N)$  invariance of the 2PI effective
action $\Gamma[\phi,\Delta]$ we  may find a WI valid in the 2PI formalism at the extremal
point of the 2PI effective action~\cite{PilaftsisTeresi}:
\begin{equation}
  \label{eq:2PI_WI}
v \, \int_x \frac{\delta^2 \Gamma[\phi,\Delta]}{\delta G_x \delta G_y}
\ +\ 2\,\frac{\delta^2 \Gamma[\phi,\Delta]}{\delta G_y \,\delta
  \Delta^{GH}_{xz}}\, \Big( \Delta_{xz}^{HH}\: -\: \Delta_{xz}^{GG} \Big)
\ =\ 0 \;. 
\end{equation}
When $\mathbb{O}(N)$-symmetric truncations
of the 2PI effective  action are considered, the approximate solutions
$\phi^i_x$  and  $\Delta^{ij}_{xy}$  satisfy \eqref{eq:2PI_WI},  which
is the WI of the 2PI formalism, but not, in general, the Goldstone theorem, which is a WI of the 1PI formalism. Unlike the latter,  the
WI~\eqref{eq:2PI_WI}  does not  protect the masslessness of
the Goldstone boson.  In
fact,        as    it  has        explicitly        been        shown
in~\cite{AmelinoCamelia,Petropoulos,Lenaghan}, the Goldstone boson $G$
is not massless in the  HF approximation, manifesting itself as a pole
at $k^2\neq 0$ in the Goldstone-boson propagator~$\Delta^{GG}(k)$. We stress here that the exact solutions obtained from the \emph{complete} 2PI effective action satisfy, instead, the Goldstone theorem,  since  the  complete  1PI and  2PI effective   actions   are   fully   equivalent   at   their   extremal
points~\cite{CJT}.

\section{Symmetry-improved 2PI Effective Action}\label{sec:SICJT}
In this section we present the symmetry-improved formalism for the 2PI effective
action $\Gamma[\phi,\Delta]$, which respects the Goldstone theorem for
any $\mathbb{O}(N)$-symmetric  truncation of~$\Gamma[\phi,\Delta]$.  

From the discussion in the previous section, it is clear that
when a given truncation of the 2PI effective action~$\Gamma_{\rm{tr}}[\phi,\Delta]$ is considered, the solution obtained extremizing $\Gamma_{\rm{tr}}[\phi,\Delta]$ with respect to
$\phi$   and   $\Delta$ fails to satisfy the Goldstone theorem
\begin{equation}
  \label{eq:WI}
v \int_x \Delta_{xy}^{-1,GG}\ =\ v \,
\Delta^{-1,GG}(k)\Big|_{k = 0} \ = 0 \; .
\end{equation}  
The symmetry-improved EoMs are obtained by imposing the 1PI WI \eqref{eq:WI} directly as a constraint for the Goldstone propagator. This can be achieved, in a variational formulation, by performing a constrained extremization of the 2PI effective action, i.e. by finding the extremum of
\begin{equation}
\widetilde \Gamma[\phi,\Delta,\ell] \ =\ \Gamma_{\rm{tr}}[\phi,\Delta] \: -\:
\ell_y^0 \, \phi^H_x \, \Delta^{-1,GG}_{xy}\; ,
\end{equation}
where $\ell_x^0$ is a Lagrange-multiplier field. In \cite{PilaftsisTeresi} it is shown that this procedure results in replacing the 2PI EoM with the 1PI one
\begin{equation}
\frac{\delta \Gamma_{\rm{tr}}[\phi, \Delta]}{\delta
  \phi^H_x}\ =\ 0 \quad \longrightarrow \quad \frac{\delta \Gamma_{\rm 1PI}[\phi]}{\delta
  \phi^H_x} = 0 \;.
\end{equation} 
The latter is explicitly given by the 1PI WI
\begin{equation}\label{eq:1PI_WI}
\frac{\delta \Gamma_{\rm 1PI}[\phi]}{\delta
  \phi^H_x} \ =\ \phi^H_x \: \Delta^{-1,GG}_{xy}[\phi ] \;.
\end{equation}
Therefore, assuming homogeneity, the symmetry-improved EoMs to solve in the SSB phase are
\begin{subequations}
 \label{eq:EoM}
\begin{align}
\frac{\delta \Gamma_{\rm{tr}}[v,\Delta]}{\delta \Delta^{ij}(k)}\ &=\ 0 \;,\label{eq:EoM_prop}\\
v\,\Delta^{-1,GG}(k)\Big|_{k=0}\ &=\ 0 \;, \label{eq:EoM_constraint}
\end{align}
\end{subequations}
namely the standard 2PI EoM for the propagators and the Goldstone theorem itself. In the  symmetric phase  of the
theory  we have $v  = 0$  and only \eqref{eq:EoM_prop} needs to be solved.

For homogeneous background field values $\phi$ away  from the minimum
$v$, i.e.~for  $\phi \neq v$, we may define a symmetry-improved effective potential $\widetilde
V_{\rm{eff}}(\phi)$ by means of the 1PI WI~\eqref{eq:1PI_WI}, seen as a differential equation defining the improved potential. We thus have the definition~\cite{PilaftsisTeresi}  
\begin{equation}
   \label{eq:WI_phi}
 - \, \frac{d \widetilde
  V_{\rm{eff}}(\phi)}{d \phi} \ \equiv \ \phi \, \Delta^{-1,GG}(k=0;\phi) \;,
\end{equation}
where $\Delta^{-1,GG}(k=0;\phi)$ is the Goldstone component of the solution of the EoM for the propagators~\eqref{eq:EoM_prop} for generic $\phi \neq v$. This definition is manifestly compatible with the EoMs at $\phi = v$ \eqref{eq:EoM}. The integral form of \eqref{eq:WI_phi} is
\begin{equation}
  \label{eq:Veff}
\widetilde{V}_{\rm{eff}}(\phi)\ =\ -\,\int_0^\phi d\phi\: \phi\,
                                   \Delta^{-1,GG}(k=0;\phi) \ +\
         \widetilde{V}_{\rm{eff}}(\phi = 0)\; . 
\end{equation}
We show in~\cite{PilaftsisTeresi} that the additive constant $\widetilde{V}_{\rm{eff}}(\phi = 0)$ can be chosen such that the formalism is 
ther\-mo\-dynamically consistent in the sense
discussed first by Baym in~\cite{Baym_2}. However,  for the purposes of this work, we do not need to determine it. We refer the interested reader to~\cite{PilaftsisTeresi} for a more detailed discussion.

\section{The Hartree-Fock Approximation}\label{sec:HF}

In  this section  we apply  the symmetry-improved  approach outlined in the previous section to  the HF
approximation  of  the 2PI  effective
action. We show  that the predicted
Goldstone boson  is massless, as it should be, and that the phase-transition  is second order already  in this  approximation. These  predictions are  in agreement with general field-theoretic properties  that hold for the full effective action of the theory, thus showing that the symmetry-improved formalism gives actually \emph{improved} predictions as compared to the standard truncated effective action.

\begin{figure}[t]
\begin{equation*}
\Gamma^{(2)}
\quad=\quad \parbox{0.63\textwidth}{\vspace{1.3em}\includegraphics[height=7.5em]{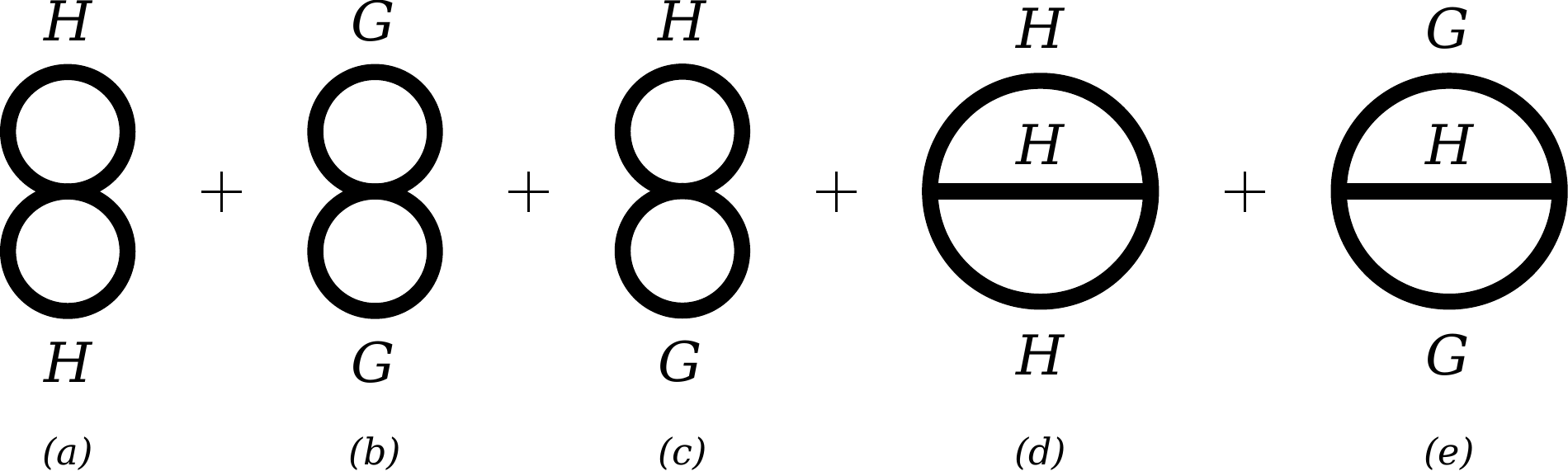}}  
\end{equation*}
\caption{Unrenormalized two-loop contributions to
  $\Gamma[\phi,\Delta]$, with thick lines denoting dressed
  propagators. The HF approximation consists of the graphs (a), (b)
  and (c) and the sunset approximation includes the graphs (d) and
  (e). \label{fig:Gamma_2}}
\end{figure}

The  HF approximation consists in retaining only  the
\emph{double-bubble}  graphs  (a)--(c)  of Fig.~\ref{fig:Gamma_2}. The unrenormalized 2PI effective action in this approximation is given by
\begin{align}\label{eq:hartree_bare}
\Gamma_{\rm{HF}}[v,\Delta^H,\Delta^G] \ &= \ \int_x
\left(\frac{m^2}{2} \, v^2 \:-\: 
\frac{\lambda}{4}\, 
v^4 \right) \: -\: \frac{i}{2} \Tr \Big(\ln \Delta^{H}\Big) \: -\:
\frac{i}{2} \Tr \Big(\ln 
\Delta^{G}\Big)  \notag\\[3pt]
&\quad+ \: \frac{i}{2} \Tr \Big(\Delta^{(0)\, -1, H}\,\Delta^H \Big) 
\: + \:  \frac{i}{2} \Tr\Big(\Delta^{(0)\, -1, G}\,\Delta^G \Big)  \notag \\[3pt] 
&\quad- \: i \, \frac{- 6 i \lambda}{8} \, i \Delta^H_{xx} \, i
\Delta^{H}_{xx} \ - \ i \frac{- 
  2 i \lambda}{4} \, i \Delta^H_{xx} \, i \Delta^{G}_{xx}  \ - \   i \, 
\frac{- 6 i \lambda}{8} \, i \Delta^G_{xx} \, i \Delta^{G}_{xx}  \;,
\end{align}
In  the  above, we  have  simplified  the  notation for  the  diagonal
propagators  as  $\Delta^H \equiv  \Delta^{HH}$  and $\Delta^G  \equiv
\Delta^{GG}$.

\subsection{Renormalization}\label{sec:HF_ren}

It  has  been shown  in~\cite{vanHees_1,Blaizot,Berges}  that the  2PI
effective action  is renormalizable  with temperature-independent  counterterms (CTs).  

To renormalize the theory, we  consider all parameters occurring in the
CJT  effective  action~\eqref{eq:hartree_bare}  to  be  bare (denoted with the subscript B):
\begin{equation}
\phi_B^i \ =  \ Z^{1/2} \, \phi^i \;,\qquad m_B^2 \ = \ Z^{-1}\big( 
m^2 \,+\, \delta m^2\big)
\;,\qquad \lambda_B \ = \ Z^{-2} \big( \lambda \,+\, \delta\lambda\big) \;,
\qquad
\Delta_B^{ij} \ = \ Z \,\Delta^{ij}\;.
\end{equation} 

In the 2PI formalism, there are two distinct 2-point operators appearing in the effective actions, corresponding to the two different derivatives
\begin{equation}
  \label{eq:dm2}
\frac{\delta^2
  \Gamma_{\rm{tr}} [\phi,\Delta]}{\delta \phi \,\delta \phi}\ ,\qquad 
\frac{\delta \Gamma_{\rm{tr}} [\phi,\Delta]}{\delta \Delta}\  ,
\end{equation}
When a generic truncation of $\Gamma[\phi,\Delta]$ is considered, these two operators are independent (including their divergences), and need to be renormalized by two different CTs~\cite{Berges}. Analogously, in the truncated 2PI formalism we have the independent 4-point functions
\begin{equation}
  \label{eq:dlambda}
\frac{\delta^4 \Gamma_{\rm{tr}} [\phi ,\Delta]}{\delta\phi\, \delta\phi\, 
\delta\phi\, \delta\phi}\ ,\qquad
\frac{\delta^3 \Gamma_{\rm{tr}} [\phi ,\Delta]}{\delta\phi \, \delta\phi \, 
\delta\Delta}\ ,\qquad
\frac{\delta^2 \Gamma_{\rm{tr}} [\phi ,\Delta]}{\delta\Delta \, \delta\Delta}\ . 
\end{equation}
Moreover, in the $\mathbb{O}(N)$ model these operators appear in two different $\mathbb{O}(N)$-invariant combinations, denoted by $A$ and $B$. It is shown in~\cite{Berges} that only the standard perturbation-theory coupling-constant CT $\delta \lambda$ is required to renormalize the higher-loop graphs of the 2PI effective action, when going beyond the HF approximation. Hence, a finite number of CTs is needed to make the effective action finite, thus guaranteeing its renormalizability.
With  the above considerations,         the        effective
action~\eqref{eq:hartree_bare} 
reads:
\begin{align}
\label{eq:hartree_ren}
\Gamma_{\rm{HF}}[v,\Delta^H,\Delta^G] \ &= \ \int_x \left(\frac{m^2 +
  \delta m_0^2}{2} \,v^2 \: -\: 
\frac{\lambda + \delta \lambda_0}{4}  \,v^4 \right)\: 
-\: \frac{i}{2} \Tr \Big(\ln\Delta^H\Big) \: -\: 
\frac{i}{2} \Tr \Big(\ln \Delta^G \Big) \notag\\[3pt]
&\quad- \: \frac{i}{2} \Tr \Big\{\Big[ Z \, \partial^2\: +\: 
\Big( 3 \lambda + \delta \lambda_1^A + 2 \delta \lambda_1^B\Big)\,v^2
- \Big( m^2 + \delta m^2_1\Big) \Big]\,\Delta^H \Big\}  \notag\\[3pt]
&\quad- \: \frac{i}{2} \Tr \Big\{\Big[ Z \, \partial^2\: +\:
  \Big(\lambda + \delta \lambda_1^A\Big) \, v^2 - \Big(m^2 + \delta m^2_1\Big)
  \Big]\,\Delta^G \Big\}  \notag\\[3pt]
&\quad- \: i \,\frac{- i \, (3\lambda+\delta \lambda_2^A + 2 \delta
  \lambda_2^B)}{4} \, i \Delta^H_{xx} \, i \Delta^{H}_{xx} \:- \:i
\,\frac{- 2 i \, (\lambda+\delta \lambda_2^A)}{4} \, i 
\Delta^H_{xx} \, 
i \Delta^{G}_{xx}  \notag\\[3pt]
&\quad -\: i \, \frac{- i \,
  (3\lambda+\delta \lambda_2^A + 2 \delta \lambda_2^B)}{4} \, i
\Delta^G_{xx} \, i \Delta^{G}_{xx} \; ,
\end{align}
where we  may set $Z=1$ at  this order of loop  expansion. 

By canceling separately divergences and subdivergences proportional to temperature-dependent terms, in~\cite{PilaftsisTeresi} we find the following CTs in the HF approximation, in the  $\overline{\rm MS}$ scheme of dimensional regularization $d=4 - 2 \epsilon$:
\begin{subequations}
  \label{eq:HF_count}
\begin{align}
\delta \lambda_2^A\ =\ \delta \lambda_1^A\ &=\  \frac{2 \lambda^2}{16
  \pi^2 \epsilon} \, \frac{3 - \displaystyle\frac{4 \lambda}{16 \pi^2 
    \epsilon}}{1 - \displaystyle\frac{6 \lambda}{16 \pi^2 \epsilon} +
  \frac{8 \lambda^2}{(16 \pi^2 \epsilon)^2}} \;, \\
\delta \lambda_2^B\ =\  \delta \lambda_1^B\ &=\ \frac{2 \lambda^2}{16
  \pi^2 \epsilon} \, \frac{1}{1 - \displaystyle\frac{2 \lambda}{16
    \pi^2 \epsilon}} \;,  \qquad\qquad
\delta m_1^2\ \, =\ \,\frac{4 \lambda m^2}{16 \pi^2 \epsilon} \,
\frac{1}{1 - \displaystyle\frac{4 \lambda}{16 \pi^2 \epsilon}}\;. 
\end{align}
\end{subequations}
The  above  $T=0$ CTs are sufficient to renormalize the EoMs for the propagators, also when thermal effects are considered.

\subsection{Thermal Phase Transition}\label{sec:phase_tran}

In the  HF approximation, the self-energies  are momentum independent.
Therefore, we may parameterize the propagators as $\Delta^{H/G}(k) =
(k^2 - M_{H/G}^2 + i \varepsilon)^{-1}$, where the effective Higgs and
Goldstone  masses,  $M^2_H$  and  $M^2_G$,  depend  only on  the
temperature $T$. 

In the symmetric phase of the theory the constraint \eqref{eq:EoM_constraint} is automatically satisfied, and the renormalized EoMs \eqref{eq:EoM_prop} take on the form
\begin{subequations}\label{eq:ren_T_symm} 
\begin{align}
M^2_H \ &= \ - m^2 \: + \: 3 \lambda  \frac{M^2_H}{16 \pi^2}  
\ln \frac{M^2_H}{2 m^2} \: + \: \lambda \frac{M^2_G}{16 \pi^2}  
\ln \frac{M^2_G}{2 m^2} \: + \: 
3\lambda \int_{\ve k} \frac{n(\omega_{\ve k}^H)}{\omega_{\ve k}^H} \: 
+\: \lambda \int_{\ve k} \frac{n(\omega_{\ve k}^G)}{\omega_{\ve k}^G}\;,\\ 
M^2_G \ &= \ - m^2 \:+\: \lambda \frac{M^2_H}{16 \pi^2} 
\ln \frac{M^2_H}{2 m^2}\:+\: 3 \lambda \frac{M^2_G}{16 \pi^2} 
\ln \frac{M^2_G}{2 m^2} \:+\: \lambda\int_{\ve k} 
\frac{n(\omega_{\ve k}^H)}{\omega_{\ve k}^H} \:+\: 3 \lambda
\int_{\ve k} \frac{n(\omega_{\ve k}^G)}{\omega_{\ve k}^G} \;,
\end{align}
\end{subequations}
where $\int_{\bf k} \equiv  \int\! d^3{\bf k}/(2 \pi)^3$, $\omega_{\ve
  k} = \sqrt{\ve {k}^2 + M^2}$ is the on-shell energy of the (quasi)particle,
and  $n(\omega)  =  (e^{\omega/T}  - 1)^{-1}$  is  the  Bose--Einstein
distribution function. In the  symmetric phase  we obtain, as expected, a  single solution with  $M_G^2 = M^2_H$.

\begin{figure}
\centering
\includegraphics[width=0.6\textwidth]{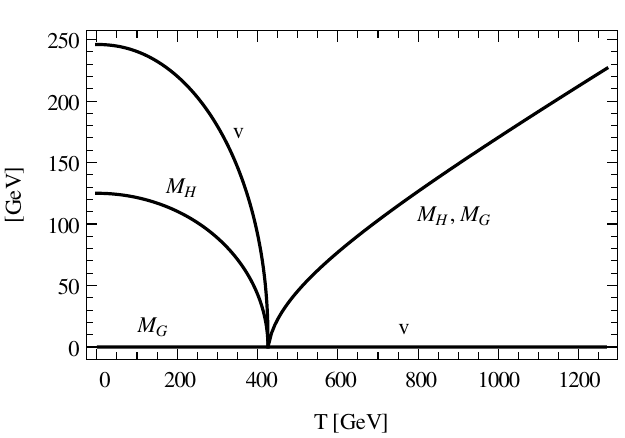}
\caption{\label{fig:hartree} The values of $M^2_H$, $M^2_G$ and  the VEV  $v$, as functions  of $T$, predicted in  the HF approximation  of the   symmetry-improved  2PI formalism.  }
\end{figure}

In the HF approximation the constraint \eqref{eq:EoM_constraint} reads
\begin{equation}
  \label{eq:constr_hartree}
v \, M_G^2 \ = \ 0 \; . 
\end{equation} 
In the SSB phase of the theory this implies $M_G^2 = 0$, yielding the EoMs~\cite{PilaftsisTeresi}
\begin{subequations}
  \label{eq:ren_T_SSB}
\begin{align}
M^2_H \ &= \ 3 \lambda v^2 \:-\: m^2 \ + \ 3 \lambda \,
\frac{M^2_H}{16 \pi^2} \, 
\ln\frac{M^2_H}{2 m^2} \ + \ 3 \lambda \int_{\ve k} \frac{n(\omega_{\ve
    k}^H)}{\omega_{\ve k}^H} \ + \ \lambda \int_{\ve k}
\frac{n(\omega_{\ve k}^G)}{\omega_{\ve k}^G}  \;,\\[3pt]
0 \ &= \ \lambda v^2 \:-\: m^2 \ + \ \lambda\,\frac{M^2_H}{16 \pi^2} \,\ln
\frac{M^2_H}{2 m^2} \ + \ \lambda \int_{\ve k} \frac{n(\omega_{\ve
    k}^H)}{\omega_{\ve k}^H} \ + \ 
3\lambda \int_{\ve k} \frac{n(\omega_{\ve k}^G)}{\omega_{\ve k}^G} \;,
\end{align}
\end{subequations}
We point out that have chosen the $\overline{\rm MS}$ mass scale $\mu$ such that the tree-level relations $M_H^2=2m^2$, $v^2=m^2/\lambda$ are satisfied at $T=0$.
The mass-gap equations
\eqref{eq:ren_T_SSB} are solved analytically to be
\begin{subequations}
\begin{align}
M^2_H \ &= \ 2 m^2 \:-\: \frac{8 \,\lambda T^2}{12} \;, \\[2mm]
M^2_G \ &= \ 0 \;, \\
v^2 \ &= \ \frac{m^2}{\lambda} \:-\: \frac{M^2_H}{16 \pi^2} \,
\ln\frac{M^2_H}{2 m^2} \:-\: 
\int_{\ve k} \frac{n(\omega_{\ve k}^H)}{\omega_{\ve k}^H} \:-\:
\frac{3 \,T^2}{12} \ . 
\end{align}
\end{subequations}
 In Fig.~\ref{fig:hartree},  we  exhibit  the  dependence of  the  squared thermal masses,  $M^2_H$ and $M^2_G$, and  the thermally-corrected VEV $v$, as  functions of the temperature  $T$.  
 The parameters of the model are  chosen  such  that  $M_H =  125~\text{GeV}$  and  $v=246~\text{GeV}$  at $T=0$. 
We observe that  we
predict   a  second-order   phase   transition at $T=T_c=\sqrt{3} v(T=0)$, already   in  the   HF
approximation,  in  agreement  with  theoretical expectations  to  all
orders.    This   is in sharp contrast  with   the  incorrect first-order
phase-transition  predicted in  the HF  approximation by  the previous
approaches~\cite{vanHees_3, Ivanov_1, Ivanov_2}.

\subsection{Symmetry-improved Effective Potential}
\begin{figure}[t]
\parbox{0.5\textwidth}{\includegraphics[height=0.23\textheight]{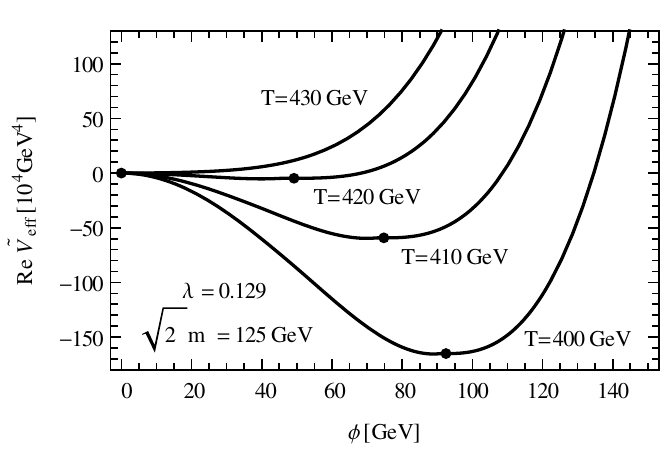}}
\parbox{0.48\textwidth}{\hspace{0pt}\includegraphics[height=0.23\textheight]{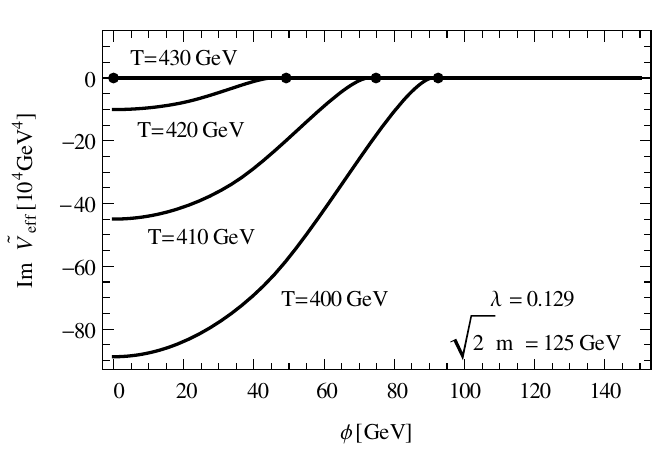}}
\caption{Symmetry-improved  HF  effective  potential  in the high-temperature approximation.  The large dots denote the
  minimum      solutions     $\phi      =     v$,      obtained     in
  Section~\ref{sec:phase_tran}.  \label{fig:V_min}}
\end{figure}

Let us now calculate the effective potential at high temperatures in the HF approximation of the symmetry-improved formalism.
Extending    the    renormalized  EoMs~\eqref{eq:ren_T_SSB} from $v \to \phi$, we~obtain
\begin{subequations}
  \label{eq:EqV}
\begin{align}
  \label{eq:EqV1}
M^2_H(\phi) \ &= \ 3 \lambda \phi^2 \,-\, m^2 \: +\: 3 \lambda \, 
\frac{M^2_H(\phi)}{16\pi^2}  
\ln \bigg(\frac{M^2_H(\phi) }{2 m^2}\bigg) \: +\: 
\lambda \,\frac{M^2_G(\phi)}{16
  \pi^2}  \ln \bigg(\frac{M^2_G(\phi) }{2 m^2}\bigg) \notag\\
\ &\quad\ +\, 3 \lambda \int_{\ve k} \frac{n[\omega_{\ve
    k}^H(\phi )]}{\omega_{\ve k}^H(\phi) } \: +\: 
\lambda\int_{\ve k} \frac{n[\omega_{\ve k}^G(\phi )]}{\omega_{\ve k}^G(\phi)} \ ,\\[2mm]
  \label{eq:EqV2}
M^2_G(\phi) \ &= \ \lambda \phi^2 \,-\, m^2 \,+\, 
\lambda\,\frac{M^2_H(\phi)}{16 \pi^2}
\ln \bigg(\frac{M^2_H(\phi)}{2 m^2}\bigg)\: +\: 
3 \lambda \,\frac{M^2_G(\phi)}{16 \pi^2} \ln\bigg(\frac{M^2_G(\phi)}{2
  m^2}\bigg) \notag\\  
\ &\quad\ +\, \lambda\int_{\ve k} \frac{n[\omega_{\ve
    k}^H(\phi)]}{\omega_{\ve k}^H(\phi )} \,+\, 3 \lambda \int_{\ve k}
\frac{n[\omega_{\ve k}^G(\phi )]}{\omega_{\ve k}^G(\phi )} \ ,\\[4mm]
  \label{eq:diffV} 
\frac{1}{\phi} \,\frac{d \widetilde V_{\rm{eff}}(\phi)}{d
  \phi} \ &= \ M^2_G(\phi)  \;.
\end{align}
\end{subequations}
The    first    two    equations are the EoMs for the propagators, for general $\phi \neq v$.  The  latter
equation~\eqref{eq:diffV}     results     from    the  definition of the symmetry-improved effective potential \eqref{eq:WI_phi},  i.e. the 1PI WI for $\phi \neq v$. Figure~\ref{fig:V_min} presents  our numerical estimates  for the high-temperature symmetry-improved   effective    potential,  as
functions  of $\phi$,  for  different temperatures~$T$~\cite{PilaftsisTeresi}. Again, a second-order phase transition is described, already in the HF approximation.  The  fact   that  $\widetilde
V_{\rm{eff}}(\phi)$ acquires  an imaginary part  when $\phi<v$ signals
the    instability    of    the    homogeneous    vacuum    in    this
region~\cite{Weinberg_effV}.

\section{Threshold Properties}\label{sec:sunset}

In this section, we study the threshold properties of the Higgs and Goldston particles, by including the contributions from the sunset diagrams (d) and  (e) in  Fig.~\ref{fig:Gamma_2}. In particular, we will  show  that the resummed Higgs-  and Goldstone-boson propagators  predicted within our
symmetry-improved   CJT  formalism   exhibit  the   correct  threshold
properties arising from on-shell  Higgs and Goldstone particles in the
loop. Therefore, we explicitly  demonstrate that the symmetry-improved approach is consistent with the optical theorem and unitarity.

One of the common approaches in the literature, when studying the issue of symmetries in the CJT formalism, is to define an additional 2-point function as
\begin{equation}
\Delta_{\rm ext}^{-1} \ \equiv \ \frac{\delta^2 \Gamma_{\rm tr}[\phi, \Delta(\phi)]}{\delta \phi \, \delta \phi} \;,
\end{equation}
where $\Delta(\phi)$ is the solution of the standard 2PI EoM for the propagator. This 2-point function is sometimes referred to as the \emph{external propagator}. It can be shown that this function satisfies the Goldstone theorem, since the second derivative (in the Goldstone direction) of any symmetric $\phi$-only functional does so~\cite{PilaftsisTeresi}. Thus, it is often claimed that this function should be considered as the true approximation to the propagator, in the CJT formalism, rather than $\Delta$. However, the problem of this approach is that what appears in the diagrammatic series of $\Gamma[\phi,\Delta]$ is the propagator $\Delta$, not the external propagator $\Delta_{\rm ext}$. The former does not satisfy the Goldstone theorem, in the standard CJT formalism, and therefore the thresholds of the particles are described incorrectly, since the Goldstone bosons propagating within quantum loops are massive. In this sense, this approach is not capable of describing the Goldstone bosons as consistently massless quantum-mechanically, i.e. within loops. From the above discussion, it is clear that studying the threshold properties described by the symmetry-improved 2PI effective action is an important step in establishing the consistency of the formalism and show further its advantages as compared to previous approaches.

\begin{figure}[t]
\begin{align*}
i\,\Delta^{-1,\,H}(p) \ &= \  
i\,{\Delta^{(0)\,-1,H}}(p)
\;\; \parbox{0.56\textwidth}{\includegraphics[height=4em]{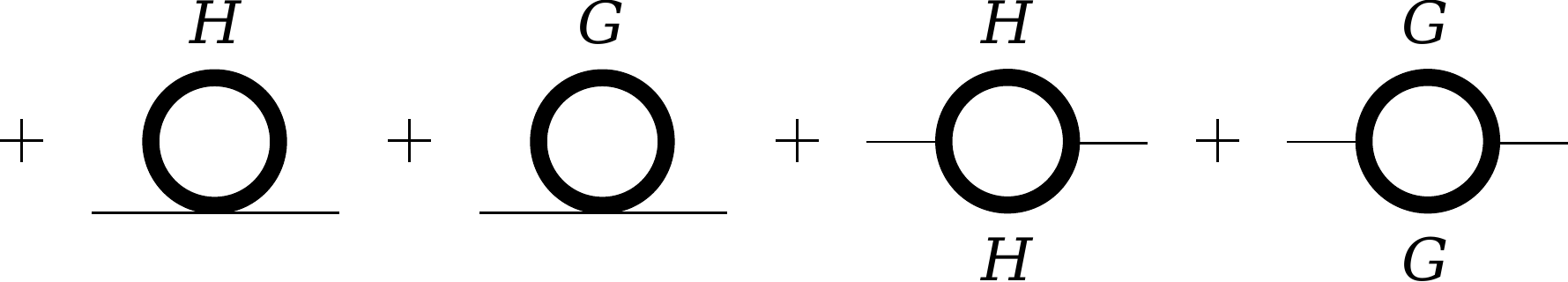}}
\\[0.5em] 
i\,\Delta^{-1,\,G}(p) \ &= \ 
\hspace{1.pt} i\,{\Delta^{(0)\,-1,G}}(p)
\;\; \parbox{0.41\textwidth}{\includegraphics[height=4em]{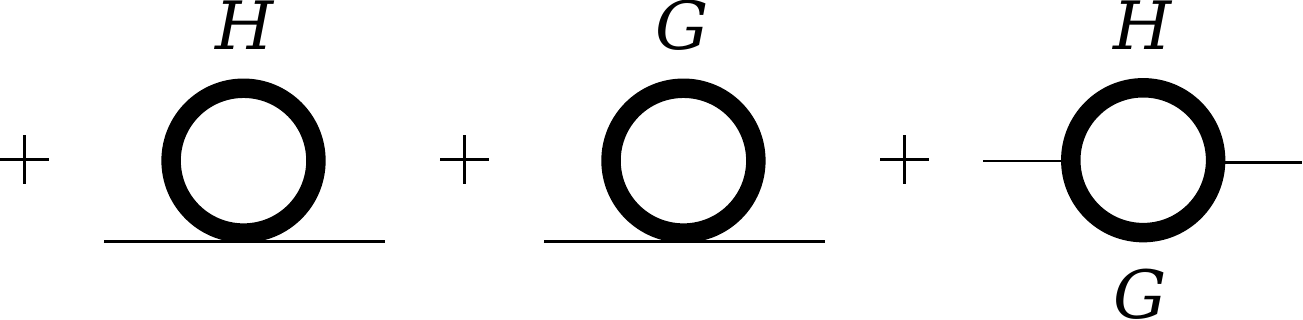}} 
\end{align*} 
\caption{Higgs and Goldstone EoMs  in   the   sunset
  approximation. \label{fig:EoMs}}
\end{figure}

As diagrammatically represented  in Fig.~\ref{fig:EoMs}, the symmetry-improved 2PI EoMs derived in the sunset approximation are given by
\begin{subequations}
  \label{eq:sun_motion}
\begin{align}
\Delta^{-1,\,H}(p) \ &= \ p^2 \:-\: (3 \lambda + \delta \lambda_1^A +
2\delta\lambda_1^B)\, v^2 \:+\: m^2 \:+\: \delta m_1^2  \:-\: (3
\lambda + \delta\lambda_2^A + 2 \delta \lambda_2^B) \, \mathcal{T}_H \notag\\[3pt] 
&\quad \:-\: (\lambda + \delta \lambda_2^A) \,\mathcal{T}_G \: + \:
\frac{1}{i} \, \frac{(- 6 i \lambda
  v)^2}{2} \, \mathcal{I}_{HH}(p) \:+\: \frac{1}{i} \, \frac{(- 2 i
  \lambda v)^2}{2} \, \mathcal{I}_{GG}(p) \;, 
\displaybreak[0]\\[9pt]
\Delta^{-1,\,G}(p) \ &= \ p^2 \:-\: (\lambda + \delta \lambda_1^A) \,
v^2 \:+\: m^2 \:+\: 
\delta m_1^2 \notag\\[3pt]
&\quad \:-\: (\lambda + \delta \lambda_2^A) \, \mathcal{T}_H
\:-\:  (3 \lambda + \delta \lambda_2^A + 2 \delta \lambda_2^B)
\, \mathcal{T}_G \:+\; \frac{1}{i} \,(- 2 i \lambda v)^2 \,
\mathcal{I}_{HG}(p) \;, 
\displaybreak[0]\\[9pt]
v \, \Delta^{-1,\,G}(0) \ &= \ 0 \;.
\end{align}
\end{subequations}
Here, we have abbreviated the loop integrals as follows:
\begin{equation}\label{eq:TaIab}
\mathcal{T}_a \ =\ \overline{\mu}^{2\epsilon} \int_k i \Delta^a(k)
\;, \qquad\qquad  
\mathcal{I}_{ab}(p) \ =\ \overline{\mu}^{2\epsilon} \int_k
i \Delta^a(k - p) \, i \Delta^b(k) \;, 
\end{equation}
where $a,b =  H,G$, $\ln\overline{\mu}^2 = \ln \mu^2  + \gamma - \ln(4
\pi)$   and   $\mu$  is   the   $\overline{\rm  MS}$   renormalization
scale. In \eqref{eq:sun_motion} we  have introduced  the  shorthand notation  $\int_k
\equiv \int\! d^4 k/(2 \pi)^4$. Details of the renormalization and  the numerical approach to solve these self-consistent nonlinear equations are given in~\cite{PilaftsisTeresi}. Here, we only outline the physical content of their solution.

\begin{figure}[t]
\parbox{0.49\textwidth}{\includegraphics[height=0.24\textheight]{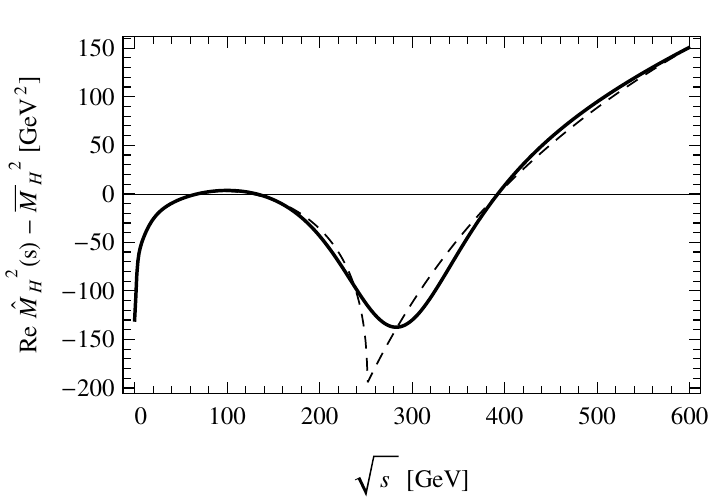}}
\parbox{0.48\textwidth}{\hspace{7pt}\includegraphics[height=0.24\textheight]{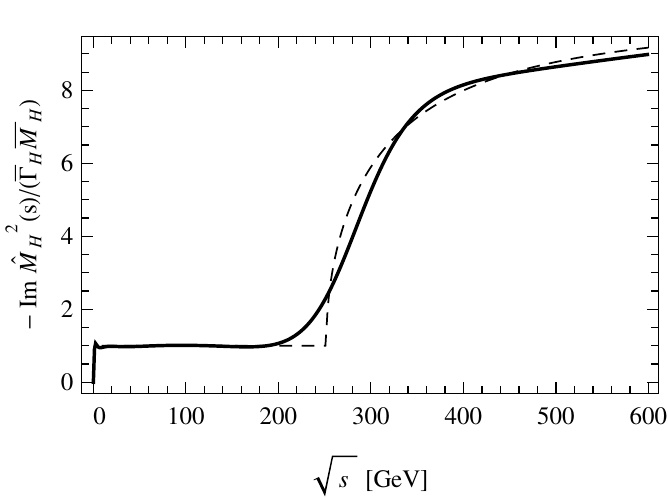}}
\caption{Numerical  solutions   for  ${\rm  Re}[\widehat{M}^2_H(s)]  -
  \overline{M}^2_H$        (left        frame)       and        ${\rm
    Im}[\widehat{M}^2_H(s)]/(\overline{\Gamma}_H\overline{M}_H)$
  (right  frame),  where $\overline{M}_H$ and $\overline{\Gamma}_H$ are the
  Higgs-boson pole  mass and width,  respectively. The dashed lines are the one-loop results in the 1PI formalism.\label{fig:H}}
\end{figure}

\begin{figure}[t]
\parbox{0.49\textwidth}{\includegraphics[height=0.24\textheight]{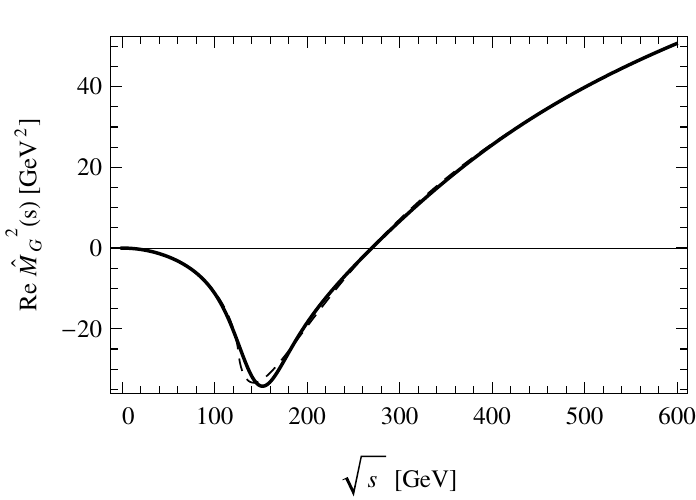}}
\parbox{0.48\textwidth}{\hspace{4pt}\includegraphics[height=0.24\textheight]{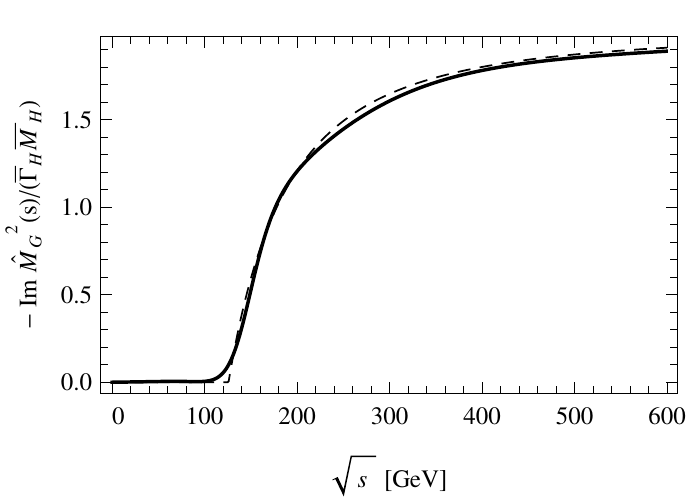}} 
\caption{Numerical solutions  for ${\rm Re}[\widehat{M}^2_G(s)]$ (left
  frame)                           and                          ${\rm
    Im}[\widehat{M}^2_G(s)]/(\overline{\Gamma}_H\overline{M}_H)$
  (right  frame).\label{fig:G}}
\end{figure}

It is useful to  define the effective
energy-dependent squared masses $\widehat{M}^2_{H/G}(s)$ as
\begin{equation}
  \label{eq:M2HGs}
\Delta^{-1,H/G} (s) \ = \ s \:-\: \widehat{M}^2_{H/G}(s) \; ,
\end{equation}
where  $s   \equiv  p^2$  is  the   Lorentz-invariant  energy-squared
parameter. In  Fig.~\ref{fig:H} and Fig.~\ref{fig:G},  we plot   the  dispersive
(real)  and  absorptive (imaginary)  Higgs- and Goldstone-boson  mass squares,  as
functions of $s$. We see
that  there is  a  non-vanishing absorptive  part ${\rm  Im}\,\widehat
M^2_H(s)$ that results  from the on-shell decay of  the Higgs particle
into two  Goldstone bosons, i.e.~$H  \to GG$.  The threshold  for this
 process is at $s = 0$, thus demonstrating that \emph{the
Goldstone  bosons in  the loop  are consistently  treated  as massless
within  our symmetry-improved  2PI formalism}.   This is, again, in sharp contrast with previous approaches in the literature, e.g. with the  results  found in~\cite{vanHees_3}  for  the absorptive part of the  external propagator, where the Goldstone boson exhibits a non-zero mass in the loop, as discussed above.

\begin{figure}[t]
\centering
\includegraphics[width=0.6\textwidth]{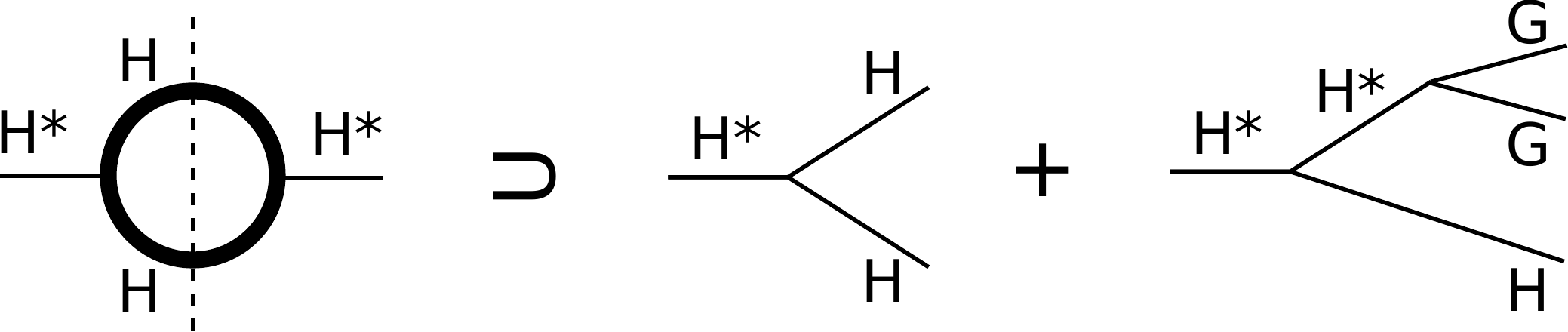}
\caption{Some of the on-shell processes whose description is included in the absorptive part of the 1-loop CJT self-energy on the LHS.\label{fig:unitarity}}
\end{figure}

The  dashed  lines   in  Figs.~\ref{fig:H}  and~\ref{fig:G}  show  the
predictions obtained at one-loop level in the 1PI
formalism. We  point out that the kinematic opening of the thresholds  is very sharp,  as opposed to the smooth thresholds predicted by  our symmetry-improved  CJT formalism. As illustrated in Fig.~\ref{fig:unitarity} this is due to the fact that \emph{the 2PI formalism automatically resums infinitely-many higher-order processes, without the need of explicitly considering them}, including also processes that take place below the 1PI threshold. 

\section{The Infrared Divergences of the Standard-Model Effective Potential}\label{sec:IR}
In this section we describe the issue, recently pointed out in~\cite{Martin:2013gka}, of the IR divergences of the SM effective potential due to the Goldstone bosons of the electroweak group. Then, by considering the scalar sector of the SM, we show how the symmetry-improved 2PI effective action can be used to study this problem, and compare the results obtained in this framework with the existing approach~\cite{Martin:2014bca,Elias-Miro:2014pca} in the literature. More technical details of the 2PI analysis will be given in a forthcoming publication~\cite{PT_new}.

\begin{figure}[t]
\centering
\begin{tabular}{rlrlrl}
$\parbox{4em}{\includegraphics[width=4em]{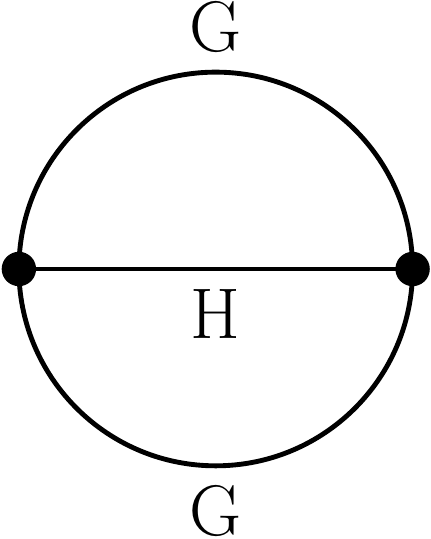}}$ & $\sim\; m^2_G \log m^2_G$ & $\parbox{6em}{\includegraphics[width=6em]{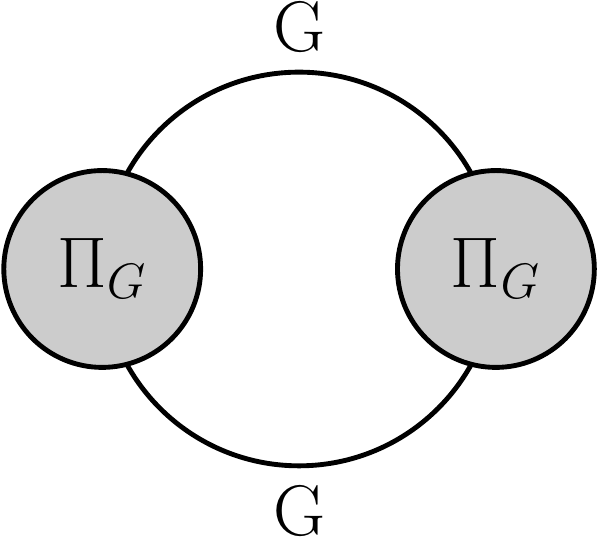}}$ & $\sim \; \log m^2_G$ & $\parbox{6em}{\includegraphics[width=6em]{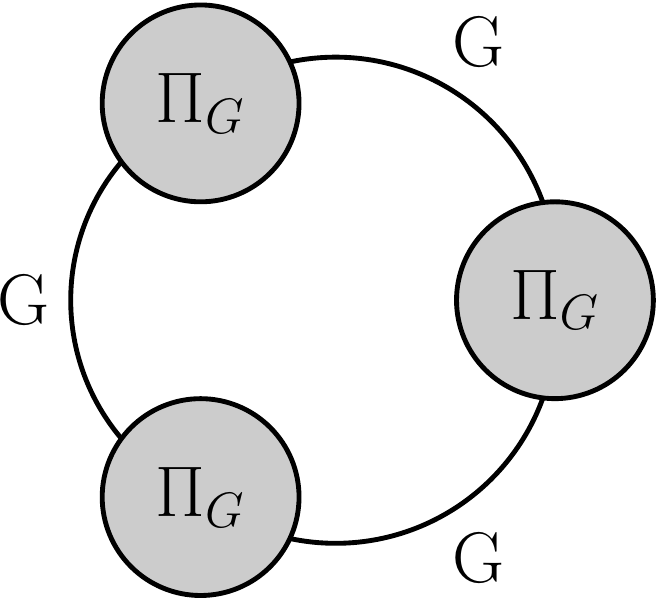}}$ & $\sim \; \displaystyle \frac{1}{m^2_G}$\\
\rule{0em}{4em}  $\displaystyle \frac{d}{d \phi} \; \parbox{4em}{\includegraphics[width=4em]{GGH}}$ & $\sim \; \log m^2_G$ & $\displaystyle \frac{d}{d \phi} \; \parbox{6em}{\includegraphics[width=6em]{GbGb}}$ & $\sim \; \displaystyle \frac{1}{m^2_G}$ & $\displaystyle \frac{d}{d \phi} \; \parbox{6em}{\includegraphics[width=6em]{GbGbGb}}$ & $\displaystyle \sim \;\frac{1}{m^4_G}$  \vspace{0.5em}
\end{tabular}
\caption{IR behaviour of Goldstone-boson ring diagrams contributing to the effective potential and its derivative. The IR divergences start at three loops for $V_{\rm eff}(\phi)$, already at two loops for its derivative $d V_{\rm eff}/ d \phi$.\label{fig:IR}}
\end{figure}

The effective potential $V_{\rm eff}$ of the SM, calculated in perturbation theory, suffers from IR divergences due to the appearance of Goldstone bosons in ring diagrams, as shown in Figure~\ref{fig:IR}. At 3 loops the divergence is logarithmic, but it becomes more and more severe with increasing loop order. Moreover, these divergences are more severe when one considers the derivative of the effective potential $\frac{d V_{\rm eff}}{d \phi}$. For the latter,  IR divergences start from 2-loop order (see Figure~\ref{fig:IR}). 

In perturbation theory these divergences appear when the tree-level Goldstone propagators become massless, i.e. at the tree-level minimum of the potential. This is instead finite, together with its derivatives, at the dressed minimum $\phi = v$. Nevertheless, this problem needs to be addressed, for the following reasons:
\begin{enumerate}
\item The effective potential $V_{\rm eff}(\phi)$ should be well-defined for all values of $\phi$, not only at its minimum $\phi = v$. Among other things, the off-shell $\phi \neq v$ effective potential governs the dynamics of the background field in inflationary scenarios.
\item At the dressed minimum $\phi = v$, the dressed masses of the Goldstone bosons vanish, and this implies that the tree-level mass $m_G^2$ is formally of the same order as the 1-loop Goldstone self-energy $\Pi_G^{(1)}$. Thus, since starting from 3-loop order the IR divergences of $\frac{d V_{\rm eff}}{d \phi}$ are of the form $1/(m_G^2)^n$, $n \geq 1$ (see Figure~\ref{fig:IR}) all \emph{these higher-loop contributions to $\frac{d V_{\rm eff}}{d \phi}$ are formally at 2-loop order}. This means that perturbation theory breaks down and these diagrams can potentially have a significant impact on 2-loop results for $\frac{d V_{\rm eff}}{d \phi}$, i.e. on the state-of-the-art threshold corrections to the VEV $v$. In view of the extreme sensitivity of the SM effective potential, extrapolated at very high energies, to the matching conditions at the electroweak scale~\cite{Bezrukov:2012sa,Degrassi:2012ry,Buttazzo:2013uya}, these issues can potentially affect the stability analyses of the SM.
\end{enumerate}
Moreover, a seemingly unrelated problem is that the tree-level mass of the Goldstone boson can be negative at the dressed minimum $\phi = v$, thus generating an unphysical imaginary part for the SM effective potential at its minimum, which does not correspond to a true instability of the homogeneous vacuum. This fact should be contrasted with the discussion in Section~\ref{sec:SICJT}, where we have shown that the symmetry-improved effective potential acquires an imaginary part only in the concave region corresponding to a physical instability. This suggests that a resummation of higher-loop diagrams is needed to address this conceptual issue as well.

Finally, we point out that the location of the IR divergence depends on the value of the gauge-fixing parameter $\xi$, but the divergence is nonetheless present in any renormalizable  $R_\xi$ gauge at the value of the field $\phi$ for which $m_G^2(\phi; \xi)=0$.

\subsection{Approximate Partial Resummation}
\label{sec:IR_pert}
Let us outline the approximate partial resummation procedure presented in~\cite{Martin:2014bca,Elias-Miro:2014pca} to address these IR issues. To facilitate an objective comparison with the 2PI approach discussed in this work, we will limit ourselves to the scalar sector of the SM, i.e. we will consider a global $\mathbb{SU}(2) \times \mathbb{U}(1)$ model with only the scalar Higgs doublet. Although this does not allow to draw quantitative conclusion for the complete SM case, this simplified model is sufficient to study this issue qualitatively and to compare the results obtained in the 2PI approach with the ones in~\cite{Martin:2014bca,Elias-Miro:2014pca}.

The partial resummation procedure consists in considering ring diagrams, as in Figure~\ref{fig:IR},  with insertions of 1-loop Goldstone self-energies $\Pi_G(k)$. One approximates these 1-loop self-energies with their zero-momentum value $\Pi_G(0)$. With this important simplification, one can resum these diagrams by replacing the 1-loop Coleman-Weinberg contribution of the Goldstone bosons with
\begin{equation}
V_{\mathrm{eff}, G}^{(1)} \ = \ \frac{3 \, m^4_G}{4 \, (16 \pi^2)} \bigg[ \log\bigg(\frac{m^2_G}{\mu^2} \bigg) - \frac{3}{2}\bigg] \quad \longrightarrow \quad  \frac{3 \, (m^2_G + \Pi_G(0))^2}{4 \, (16 \pi^2)}  \bigg[ \log\bigg(\frac{m^2_G + \Pi_G(0)}{\mu^2} \bigg) - \frac{3}{2}\bigg]\;.
\end{equation}
However, it can be shown that the derivative of this resummed term is still divergent. This problem can be solved by limiting $\Pi_G(0)$ to contain only the terms not proportional to $m_G^2$, i.e. by replacing
\begin{equation}
\Pi_G(0) \quad \longrightarrow \quad  \Pi_g \ \equiv \ \Pi_G(0) - \frac{3 \lambda}{(16 \pi^2)} \, m^2_G \Big(\log(m^2_G/\mu^2) - 1\Big) \;.
\end{equation}
Notice that the subtracted term does not correspond to the contribution of a given diagram, but it is contained in the contribution of both the Goldstone tadpole integral and the Higgs-Goldstone sunset diagram. Finally, one needs to subtract from $V_{\rm eff}$ the diagrams that would be double-counted otherwise. In conclusion, one adds to the effective potential the term~\cite{Martin:2014bca,Elias-Miro:2014pca}:
\begin{equation}
V_{\mathrm{eff}, G}^{(resum)} \ \equiv \ \frac{3 \, (m^2_G + \Pi_g)^2}{4 \, (16 \pi^2)}  \bigg[ \log\bigg(\frac{m^2_G + \Pi_g}{\mu^2} \bigg) - \frac{3}{2}\bigg] \ - \ V_{\mathrm{eff}, G}^{(d.c.)}\;, 
\end{equation}
where $V_{\mathrm{eff}, G}^{(d.c.)}$ is the contribution of the double-counted diagrams, as discussed above.

\begin{figure}[t]
\centering
\includegraphics[width=0.9\textwidth]{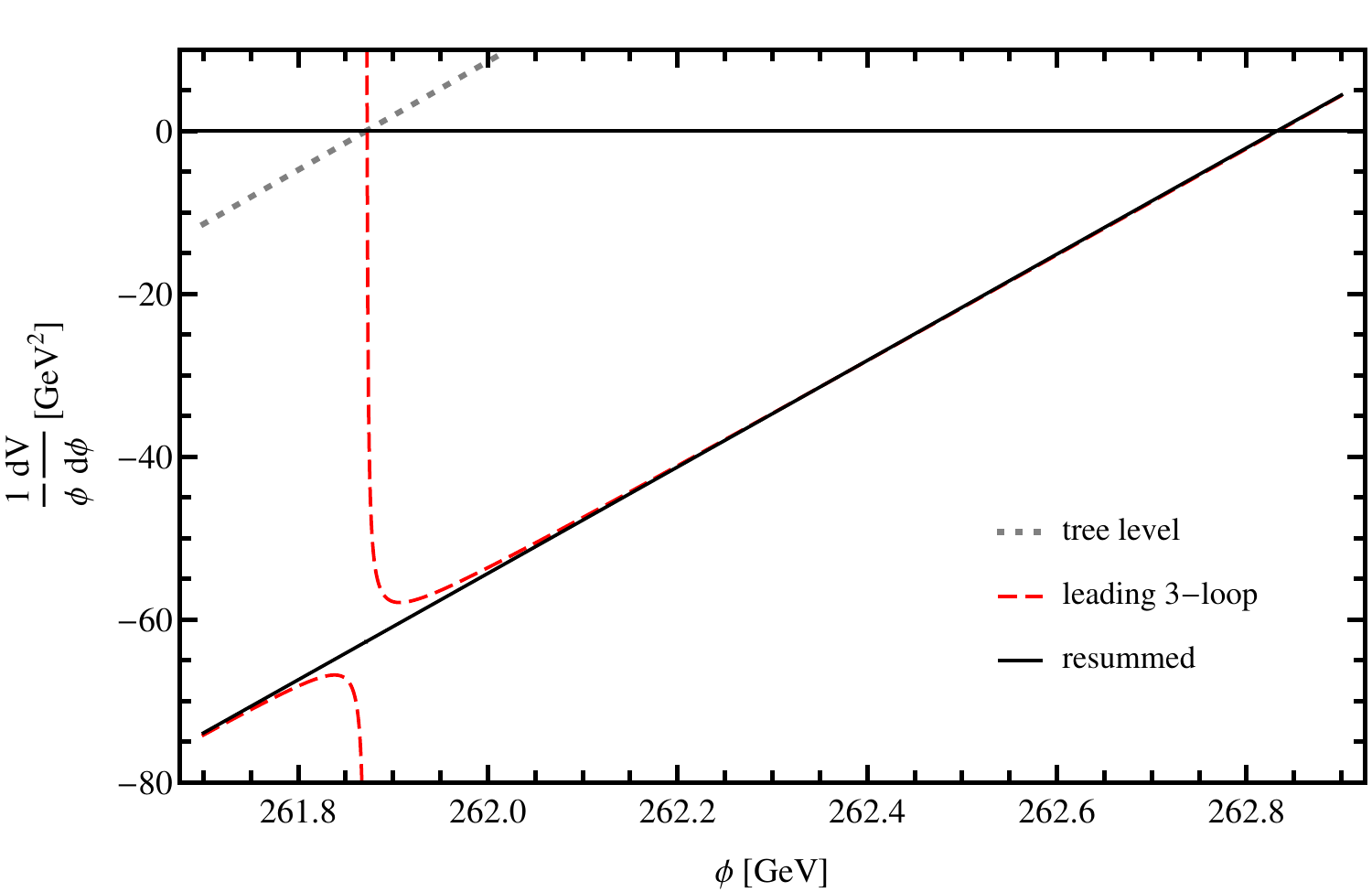}
\caption{IR divergence of the derivative of the effective potential, as calculated in perturbation theory at 3-loop order, considering only the scalar sector of the SM. The gray dotted line is the tree-level contribution, the red dashed line the leading 3-loop one, whereas the black continuous line is the result of the approximate partial resummation procedure developed in \cite{Martin:2014bca, Elias-Miro:2014pca}.\label{fig:IR_pert}}
\end{figure}

As we show in Figure~\ref{fig:IR_pert} for the scalar sector of the SM, this procedure effectively resums the IR divergences present in $d V_{\rm eff}/ d \phi$ at two- and three-loop orders. We include only the 3-loop contributions coming from the Goldstone ring diagrams in Figure~\ref{fig:IR}, since they are proportional to $1/m^2_G$ and so responsible for the IR divergence. Moreover, as we discussed above, their order gets formally lowered at the dressed minimum.

\subsection{2PI Approach to the Resummation of IR Divergences}\label{sec:IR_2PI}

We now study these issues by means of the symmetry-improved 2PI formalism. This provides a more complete resummation, as compared to the approach outlined above, for several reasons. First, it is a first-principle approach, and no ad-hoc subtraction is needed. Second, as we are going to show below, the 2PI approach takes into account more topologies and does not necessitate to neglect the momentum dependence of the self-energy insertions that are resummed. Moreover, as we showed in the previous section, the threshold properties are correctly described within the symmetry-improved formalism. In particular, the Goldstone propagator in the formalism is massless at the dressed minimum of the potential and nevertheless the effective potential will be shown to be free of IR pathologies.

\begin{figure}[t]
\centering
\begin{align*}
\Delta^{-1}(\phi) \quad &= \quad   {\Delta^{(0)\,-1}(\phi)} \quad + \quad \parbox{3.5em}{\includegraphics[width=3.5em]{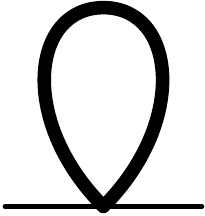}} \quad + \quad \parbox{5em}{\includegraphics[width=5em]{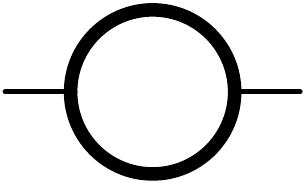}}\\[9pt]
 &\qquad + \quad \Bigg[ \parbox{5em}{\includegraphics[width=5em]{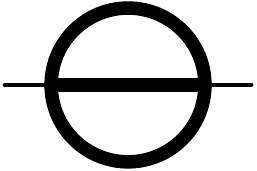}} \; + \; \parbox{4.5em}{\includegraphics[width=4.5em]{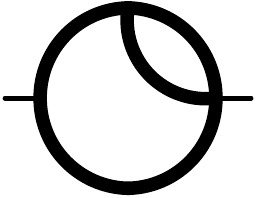}} \; + \; \parbox{4.5em}{\includegraphics[width=4.5em]{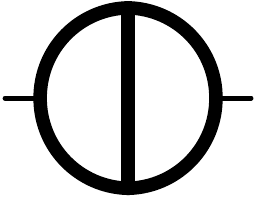}} \; + \; \parbox{7em}{\includegraphics[width=7em]{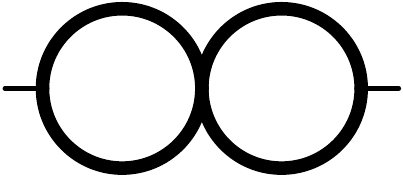}} \; \Bigg]_{\Delta \approx \Delta_0(\phi)}
\end{align*}
\caption{Diagrammatic representation of the approximation scheme for the EoMs in Section~\ref{sec:IR}.\label{fig:EoM}}
\end{figure}

\begin{figure}
\centering
\includegraphics[width=0.95\textwidth]{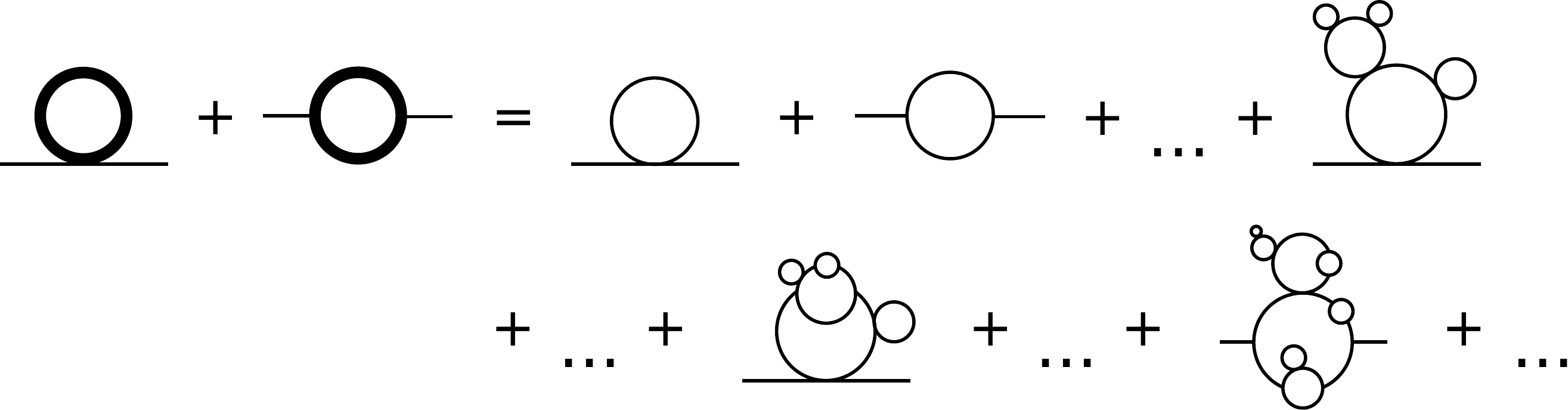}
\caption{Some of the topologies implicitly resummed by the 1-loop 2PI self-energies in the first line of Figure~\ref{fig:EoM}.\label{fig:resum_1loop}}
\end{figure}

\begin{figure}
\centering
\includegraphics[width=0.85\textwidth]{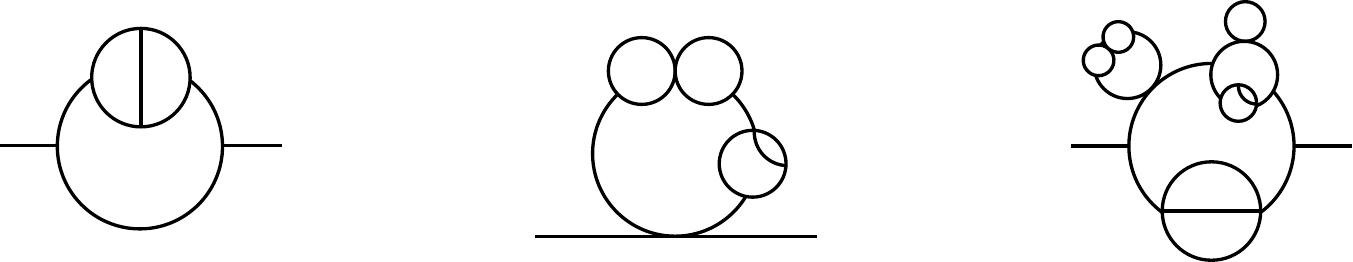}
\caption{Some examples of the topologies resummed by including the 2-loop 2PI self-energies in the second line of Figure~\ref{fig:EoM}. Notice that the propagators belonging to 2-loop 2PI topologies do not get dressed, because of the approximation $\Delta(\phi) \approx \Delta^{(0)}(\phi)$ used, for these, in Figure~\ref{fig:EoM}.\label{fig:resum_2loop}}
\end{figure}

We consider the global $\mathbb{SU}(2) \times \mathbb{U}(1)$ scalar model. By including, in the EoMs, the 1-loop (HF + sunset) CJT diagrams shown in the first line of Figure~\ref{fig:EoM}   we actually resum, automatically, a large class of diagrams, as depicted in Figure~\ref{fig:resum_1loop}. Moreover, in order to be able to compare our results with the perturbative 2-loop calculation, we include the 2-loop CJT diagrams in the second line of Figure~\ref{fig:EoM}. However, to simplify the treatment, we approximate the propagators appearing in these diagrams as the tree-level ones $\Delta(\phi) \approx \Delta^{(0)}(\phi)$. In this way, we take into account the full contribution of 2-loop topologies and, in addition, resum diagrams as the ones shown in Figure~\ref{fig:resum_2loop}. Therefore, a much larger class of diagrams is included, as compared to the method outlined in Section~\ref{sec:IR_pert}, and the momentum dependence of the resummed insertions is also retained.

In the symmetry-improved 2PI approach, the IR divergences are absent by construction: IR divergences can be present only when two or more Goldstone propagators carry the same momentum, as in the ring diagrams shown in Figure~\ref{fig:IR}. In other words, the IR pathologies originate from chains of Goldstone lines with self-energy insertions between them. However, such topologies are necessarily \emph{2-particle-reducible} and thus do not appear in the diagrammatic series of $\Gamma[\phi,\Delta]$. Therefore, the resummation of IR divergences is achieved automatically by the construction of the 2PI effective action. 

\begin{figure}[t]
\centering
\includegraphics[width=0.9\textwidth]{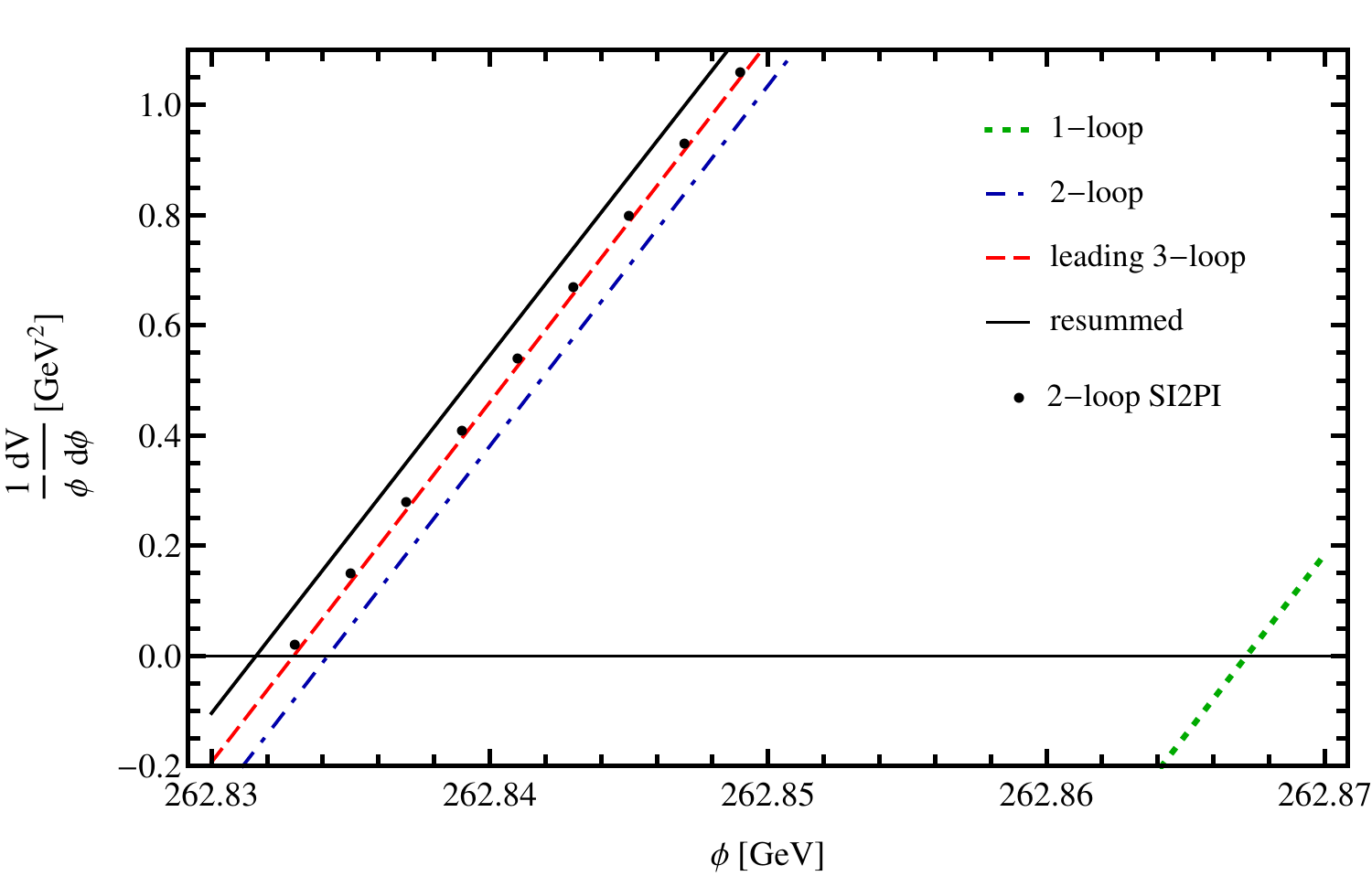}
\caption{The derivative of the effective potential, near its dressed minimum. We show results from 1-loop (green dotted line), 2-loop (blue dash-dotted line) and leading 3-loop (red dashed line) perturbation theory. The black continuous line is the result of the approximate partial resummation procedure discussed in Section~\ref{sec:IR_pert}. The black dots are the results obtained from the symmetry-improved 2PI effective action.\label{fig:IR_SICJT}}
\end{figure}

The EoMs can be easily obtained by generalizing \eqref{eq:sun_motion}  and including the 2-loop self-energies, in which $\Delta(\phi) \approx \Delta^{(0)}(\phi)$. Their explicit form is given in~\ref{app:EoMs}.  The numerical solution, in the vicinity of the dressed minimum $\phi=v$, is plotted in Figure~\ref{fig:IR_SICJT}. The black dots represent the numerical solution obtained in our approach. It is apparent that the results from the partial resummation outlined in Section~\ref{sec:IR_pert} do not reproduce the ones obtained in the more complete 2PI resummation. We have checked explicitly that, expanding the EoMs at 2-loop order, we reproduce numerically the results coming from 2-loop perturbation theory. Notice also that the 3-loop result is larger than what one would naively expect (roughly about $\lambda/16 \pi^2$ times the 2-loop one), because of the breakdown of perturbation theory, as discussed at the beginning of this section. In Figure~\ref{fig:IR_SICJT} the symmetry-improved 2PI solution appears to be similar in size to the leading 3-loop result, potentially indicating that the effect of the resummation is small near the dressed minimum. However, this is a coincidence and depends on the class of topologies of graphs considered here. A more detailed study will be given in~\cite{PT_new}.

\section{Conclusions}\label{sec:conclusions}
The 2PI effective action provides a powerful theoretical tool to consistently resum infinite series of perturbation-theory diagrams of different topologies. However, its  loopwise  expansion  introduces
residual  violations  of possible  global  symmetries by  higher-order
terms. In the case of global symmetries, this leads to the appearance of massive Goldstone  bosons in the spontaneously  broken phase  of  the  theory. In this work we have reviewed the symmetry-improved CJT formalism, developed in~\cite{PilaftsisTeresi} for   consistently   encoding   global
symmetries in loopwise expansions of the 2PI effective
action. 

We have demonstrated, in a simple $\mathbb{O}(2)$  scalar model, the key field-theoretical properties of the formalism. In detail, we have shown that the Goldstone bosons are described as massless, also within quantum loops, thus providing a consistent description of the threshold properties of the Higgs and Goldstone particles. Moreover, the thermal phase transition is at second order already in the HF approximation. Thus, the behaviour expected for the full theory is recovered already in the first non-trivial approximation, contrary to other approaches in the literature.

Thanks to the satisfactory field-theoretical properties of the formalism, we have applied the symmetry-improved 2PI effective action to the study of the IR divergences of the SM effective potential due to the electroweak Goldstone bosons. By limiting ourselves to the scalar sector of the electroweak group, we have confirmed that these IR divergences are indeed an artifact of perturbation theory, and are absent in the symmetry-improved 2PI formalism, as it should be. We have compared quantitatively our results with the predictions given by the existing approach in the literature, consisting in an approximate partial resummation of Goldstone-boson ring diagrams. We have shown that the latter, whilst correctly resumming qualitatively the IR divergences, does not reproduce quantitatively, at least in the simplified scalar model considered here, the results of the more complete 2PI resummation, near the dressed minimum of the effective potential. 

In view of the extreme sensitivity of the SM effective potential, extrapolated to very high energies, to the matching conditions at the electroweak scale, it is important to assess these issues in a more realistic way, going beyond the scalar sector of the SM. A detailed study of this matter, including the quantitatively most important contributions, will be given in a forthcoming publication.

\ack

The work  of A.P. is supported  by the
Lancaster-Manchester-Sheffield  Consortium   for  Fundamental  Physics
under  STFC   grant  ST/L000520/1.  The work of D.T. is supported by a fellowship of the EPS  Faculty of  the  University  of Manchester.

\appendix

\section{Equations of Motion for the $\mathbb{SU}(2) \times \mathbb{U}(1)$ Scalar Model}\label{app:EoMs}
In this appendix we give the renormalized EoMs used in Section~\ref{sec:IR_2PI} to study the IR divergences of the effective potential. Since the scalar $\mathbb{SU}(2)_L \times \mathbb{U}(1)$ model  automatically possesses the custodial symmetry $\mathbb{SU}(2)_L \times \mathbb{SU}(2)_R \sim \mathbb{O}(4)$, the HF and sunset contributions to the EoMs are easily inferred from the $\mathbb{O}(2)$ case (see~\cite{PilaftsisTeresi}). We find, in Euclidean space:
\begin{subequations}
\begin{align}
\Delta^{-1,\,H}(p; \phi) \ &= \ p^2 \:+\: 3 \lambda \phi^2 \,-\, m^2  \:+\: 3
\lambda  \, \mathcal{T}_H^{\rm ren} \:+\: 3 \lambda \,\mathcal{T}_G^{\rm ren} \: - \:
 18\lambda^2\phi^2 \, \mathcal{I}_{HH}^{\rm ren}(p) \: - \:
 6\lambda^2\phi^2 \, \mathcal{I}_{GG}^{\rm ren}(p) \notag\\
 & \quad + \:\Pi^{\mathrm{2PI}, (2)}_H(p;\phi)   \;, \\[3pt]
\Delta^{-1,\,G}(p; \phi) \ &= \ p^2 \:+\: \lambda \phi^2 \,-\, m^2  \:+\: \lambda  \, \mathcal{T}_H^{\rm ren} \:+\: 5 \lambda \,\mathcal{T}_G^{\rm ren} \: - \:
 4\lambda^2\phi^2 \, \mathcal{I}_{GH}^{\rm ren}(p) \: + \:\Pi^{\mathrm{2PI}, (2)}_G(p;\phi)   \;, \\[3pt]
\frac{1}{\phi} \,\frac{d \widetilde V_{\rm{eff}}(\phi)}{d
  \phi} \ &= \ \Delta^{-1,\,G}(0; \phi) \,,
\end{align}
\end{subequations}
where the $\overline{\rm MS}$ renormalized sunset and tadpole integrals are given by~\cite{PilaftsisTeresi}
\begin{align}
\mathcal{I}_{ab}^{\rm ren}(p) \ &\equiv \ \int_k \bigg( \Delta^a(k-p)\, \Delta^b(k) \:-\: \frac{1}{(k^2 + \mu^2)^2}\bigg) \;,\\
\mathcal{T}_a^{\rm ren} \ &\equiv \ \int_k \bigg[ \Delta^a(k) \:-\: \frac{1}{k^2 + \mu^2} \: - \: \frac{1}{(k^2 + \mu^2)^2} \bigg(\mu^2 - M^2_a + \frac{\nu_a \, \lambda^2 \phi^2}{16 \pi^2} B(p;\mu^2,\mu^2) \bigg)  \bigg] \notag\\
& \quad - \: \frac{\mu^2}{16 \pi^2} \: + \: \frac{\nu_a \, \lambda^2 \phi^2}{(16 \pi^2)^2} \,\frac{\eta}{2} \;, \label{eq:ren_T}
\end{align}
where 
$\nu_H = 24$, $\nu_G = 4$ and we have introduced
\begin{align}
&\frac{1}{16 \pi^2} \, B(p; \mu^2, \mu^2) \ \equiv \ \mathcal{I}_{00}^{\rm ren}(p)\;, \qquad \text{with} \quad \Delta^0(k) \equiv \frac{1}{k^2 + \mu^2}\;, \\
&\eta \ = \ 1 \; - \; \frac{4 i}{\sqrt{3}} \left( \mathrm{Li}_2\frac{1 - i \sqrt{3}}{2} - \frac{\pi^2}{36}\right) \ \simeq \ -1.34391\;.
\end{align}
Here, we seize the opportunity to eliminate an error in the value of $\eta$ reported in~\cite{PilaftsisTeresi}.
The terms in the second line of~\eqref{eq:ren_T} are the $\overline{\rm MS}$ finite part of the CT integrals used to renormalize $\mathcal{T}_a$ by the method discussed in~\cite{PilaftsisTeresi}. Their subtraction is needed to guarantee that the renormalization scheme used here matches the standard $\overline{\rm MS}$ one in perturbation theory. Finally, the 2-loop 2PI self-energies $\Pi^{\mathrm{2PI}, (2)}_{H,G}(p;\phi) $ are calculated in perturbation theory by standard techniques~\cite{Martin:2003it, Martin:2003qz}. We approximate them by their zero-momentum value $\Pi^{\mathrm{2PI}, (2)}_{H,G}(\phi)$, since the error introduced in this way is expected to be negligible for the purposes of this work. Adopting the compact notation used in~\cite{Martin:2003it}, in terms of the functions defined there we find
\begin{subequations}
\begin{align}
(16 \pi^2)^2\, \Pi^{\mathrm{2PI}, (2)}_{H}(\phi) \ &= \ 54 \lambda^3 \phi^2 \, \overline{\ln}^2 H \: + \: 36 \lambda^3 \phi^2 \, \overline{\ln}\, H \overline{\ln}\, G \:+\: 30 \lambda^3 \phi^2 \, \overline{\ln}^2 G \notag \\
&- \ 6 \lambda^2 \, I(H,H,H) \:- \: 6 \lambda^2 \, I(H,G,G) \notag\\
&- \ 216 \lambda^3 \phi^2 \, I(H',H,H) \:-\: 72 \lambda^3 \phi^2 \, I(H',G,G) \: - \: 24 \lambda^3 \phi^2 \, I(G',G,H) \notag\\
&- \ 648 \lambda^4 \phi^4 \, I(H',H',H) \:-\: 144 \lambda^4 \phi^4 \, I(H',G',G) \: - \: 24 \lambda^4 \phi^4 \, I(G',G',H) \;, \displaybreak[0]\\
(16 \pi^2)^2\, \Pi^{\mathrm{2PI}, (2)}_{G}(\phi) \ &= \ 8 \lambda^3 \phi^2 \, B(G,H)^2  \: - \: 24 \lambda^2 I(H,H,H) \:+\: 22 \lambda^2 I(G,H,H) \notag \\
& -\: 16 \lambda^2 I(G,G,H) \:+\: 6 \lambda^2 I(G,G,G) \;,
\end{align}
\end{subequations}
where all functions have to be evaluated at zero momentum.

\section*{References}


\end{document}